\documentclass[aps,prd,amssymb,amsmath,nofootinbib]{revtex4}
\usepackage[dvips]{graphicx}
\begin{document}
\title{Measures and metrics in uniform gravitational fields\footnote{This paper
is the continuation of a previous work \cite{albe} of the same author and
contains most part of it, but some definitions are different. From section
\ref{spacedil} different new calculations are added or modified and new results
are derived; section \ref{desloge} contains only new material.}}
\author{Marco Alberici}
\affiliation{Dipartimento di Fisica, Universit\`a di Bologna Via Irnerio~46,
40126 Bologna, Italy.}
\email[]{alberici@bo.infn.it}
\altaffiliation{I.N.F.N. member, Sezione di Bologna.}
\date{\today}
\begin{abstract}
A partially alternative derivation of the expression for the time dilation
effect in a uniform static gravitational field is obtained by means of a
thought experiment in which rates of clocks at rest at different heights are
compared using as reference a clock bound to a free falling reference system
(FFRS). Derivations along these lines have already been proposed, but generally
introducing some shortcut in order to make the presentation elementary. The
treatment is here exact: the clocks whose rates one wishes to compare are let
to describe their world lines (Rindler's hyperbolae) with respect to the FFRS,
and the result is obtained by comparing their lengths in space-time. The
exercise may nonetheless prove pedagogically instructive insofar as it shows
that the some results of General Relativity (GR) can be obtained in terms of
physical and geometrical reasoning without having recourse to the general
formalism. The corresponding GR metric is derived, to the purpose of making a
comparison with solutions of Einstein field equation and with other metrics.
For this reasons this paper also compels to deal with a few subtle points
inherent in the very foundations of GR.
\end{abstract}
\maketitle
\newcommand{\be}{\begin{eqnarray}}
\newcommand{\ee}{\end{eqnarray}}
\section{Introduction}
It is possible to show that the GR redshift formula for a uniform gravitational
field $\Delta \nu /\nu =gh/c^2$ it is an exact consequence of the EP, and its
derivation does not require GR formalism \cite{price}. As is well known, this
formula was experimentally verified for the first time in 1960 by Pound and
Rebka \cite{pound}. The height difference ($22.5m$) used in their experiment
was small compared to the gravitational field strength variations. For this
reason their experiment had the purpose to verify only the first order effects
predicted from GR that, for time dilation, bring to the above relation.
In experiments involving stronger gravitational field variations it is
necessary to use the exact Schwarzschild solution of GR, and to take into
consideration the earth rotation effects.
\par GR is a well-established theory, but it very often happens that
its applications to some specific case involves subtle points on which the
agreement is not general: as a result, the conclusions of the analysis reported
are often not univocal. This is the case, in particular, for the object of the
present article, namely redshift in a uniform gravitational field. In
Weinberg's treatise \cite{wein}, for instance, the gravitational redshift
effect is thus commented: ``For a uniform gravitational field, this result could
be derived directly from the Principle of Equivalence, without introducing a
metric or affine connection''. We have in fact already recalled that such
derivations are possible and in an exact way \cite{price}. Independently of
this conclusion, one may always introduce transformations between reference
frames at rest with respect to the matter generating the field and FFRS which,
in the case of a uniform field, are supposed to become extended, or global. The
most natural way to describe such a transformation are Rindler hyperbolae. But
Rindler himself \cite{rindler}, at the end of the paragraph of his book in
which they are introduced and analysed, and after discussing the fact that SR
can deal in a proper way with accelerated frames, states: ``A uniform
gravitational field can not be constructed by this method''. About 15 years ago,
E.A. Desloge \cite{des1,des2,des3,des4}, in a sequence of four articles,
carried out a systematic investigation on uniformly accelerated frames and on
their connection with gravitation. In agreement with Rindler's statement
Desloge proved that there is no exact equivalence between a frame uniformly
accelerated and a frame at rest in a uniform gravitational fields \cite{des3}.
By contrast, only three years later
% on this journal,
E. Fabri \cite{fabri} derived the expression of the time delay in a uniform
gravitational field by explicit use of a FFRF, from which the world lines of
objects at rest with the masses that generate the field are seen as Rindler
hyperbolae. As far I know, there are not other papers approaching the problem
in this way, except the Mould treatise \cite{mould}. The related metric is
instead used in important books and many paper (for example
\cite{misner0,vallis}).
\par In order to see whether a reconciliation between these apparently
conflicting views can be obtained, I tried to set up an ideal experiment in
which the rates of two standard clocks \footnote{We will not come back here to
the meaning to be attributed to expressions such as `Standard Clock' or
`Standard Rods', as the question has already been discussed in many books and
papers and also in one of Desloge's articles. We will just compare the proper
time elapsed in the two RF.} sitting at different heights in a uniform and
static gravitational field are compared. The measuring process as seen from a
FFRF, it is carefully analysed, following the approach based on Rindler
hyperbolae. We will show that in this case the associated field it is not
uniform if measured from an observer at rest with the FFRF, but that it appears
to be uniform when its strength is measured using radar pulses generated and
received from an observer at rest with the source generating the field. The
metric derived with the approach based on Rindler hyperbolae (we will call it
Rindler's metric) is a solution of the Einstein field equation for the flat
space ($R_{\mu \nu \sigma \rho} =0$), and it is also the first order
approximation of the Schwarzschild metric (section \ref {metric}). At the end
of the paper (section \ref{desloge}) it will be proved that Desloge's metric
\cite{des3} (rigid FFRF describing a geodetic with uniform proper acceleration)
can also be derived from an approach based on Rindler hyperbolae, but with a
different physical interpretation, that does not seems to be in agreement with
the EP. For this reasons it is author opinion that, for uniform fields
description, the Rindler's metric it is the most useful one.
\section{Free-falling reference systems}
\label{acc}
EP is the basic idea from which the bulk of GR theory was developed. In our
applications we will not deal with charged bodies and with non-gravitational
forces in general. We are therefore entitled to overlook differences between
weak and strong EP, differences that are present only when non-gravitational
interactions are considered. We just consider the EP in this form:
\begin{quote}
{\em A body in free fall does not feel any acceleration, and behaves locally as
an inertial reference system of SR.}
\end{quote}
We know that physical gravitational fields are not uniform, and there always
exist tidal forces that oblige to consider the validity of the EP only locally
in space and time. The ideal case of a uniform and static field is however
worth considering for its conceptual interest. In this ideal case FFRS, which
in strictly physical situations are local inertial systems, should become
non-local.

\par EP invites to consider FFR systems the proper inertial systems.
Let us then try to build a non-local FFRF, denoted as $K$, in a 2-dimensional
space-time universe and study the world line of a particle at rest with the
masses that generate a uniform gravitational field $g$.  Since this world line
is considered with respect to the FFRF, we will associate to the particle a
(local) frame $\Sigma^-$. It is easily proved \cite{rindler} that the world
line of a body falling along the vertical direction $x$ with constant proper
acceleration $g$ is described by the hyperbola of equation:
\be \label{eq.rin} x^{2}-c^2t^{2}=\frac{c^4}{g^{2}}. \ee
Since we are considering  acceleration with respect to $K$, it is material
points at rest in  $\Sigma^-$ that will describe an hyperbola of
Eq.~(\ref{eq.rin}) in the Minkowski space-time of $K$ (with an acceleration
pointing in the direction opposite to gravity). On the other hand, the observer
sitting in $\Sigma^-$ will see material points at rest in $K$ describe more
complicated world lines. They have been explicitly calculated \cite {hamilton},
and turn out indeed to be very different from those described by
Eq.~(\ref{eq.rin}) (on the same subject see \cite{ror}). Fortunately we do not
need to deal with them in order to solve our problem. Proper acceleration $g$
(acceleration relative to instantaneous rest frame) can be measured from an
observer at rest with the FFRF in $x_0$ or from an observer at rest with
$\Sigma^-$. At time $t=0$ the two observers have the same instantaneous local
rest frame and they measure the same local acceleration $g$ (an observer in $\Sigma$
measures a force acting on him, while one in FFRF deduces its value from $\Sigma$
trajectory).
\par In the following $x$ will be called `coordinate position' and $t$
`coordinate time', in order to distinguish them from proper coordinate and
proper time measured by $\Sigma^-$. As usual $c$ is the speed of light, but for
sake of simplicity in the following we will adopt units with $c=1$. The space
coordinate at time $t=0$ is $x_0=1/g$, so that Eq.~(\ref{eq.rin}) becomes
\be \label{eq.rin2} x^{2}-t^{2}=x_0^2. \ee
It is now necessary to require that the radar distance between observers
sitting at different height does not depend on proper time. By this request one
find that a second RF $\Sigma ^+$ which, at $t=0$, lies higher with respect
$\Sigma^-$ by the amount $h$ as measured by $K$, must be described by the
following hyperbola \cite{rindler}
\be \label{eq.rioo} x^{2}-t^{2}={(x_0+h)}^{2}, \ee
were now the acceleration $g$ depend on the coordinate. We will use $g^-$ to
denote the acceleration in $x_0$ and $g^+$ that in $x_0+h$ \footnote{It is
easily proven that $g^-$ and $g^+$ are in the following relation:
\be \label{eq.grel} g^+=\frac{g^-}{1+g^-h}.\ee}
\be \label{eq.g} g^+=\frac{1}{x_0+h} \qquad \qquad g^-=\frac{1}{x_0},\ee
\par The fact that the acceleration depends on the coordinate value, seems to
contradict the supposed uniformity of the field. Nevertheless on this problem
G.'t Hooft\cite{tooft} wrote: ``The gravitational field strength felt locally
\dots is inversely proportional to distance \dots. Even tough our field is
constant in the transverse direction and with time, it decrease with height''.
We will discuss again this fact in subsection \ref{gravityacc}, but it worth
now to underline that requiring the same acceleration in $\Sigma^-$ and
$\Sigma^+$ generates unsolvable problems. One should in fact identify
$\Sigma^+$ with a spatial translation of $\Sigma^-$:
\be \label{eq.trasl} (x- h)^{2}-t^{2}=x_0^2, \ee
\begin{figure}[hbp]
\begin{center}
\includegraphics[scale=0.20 ]{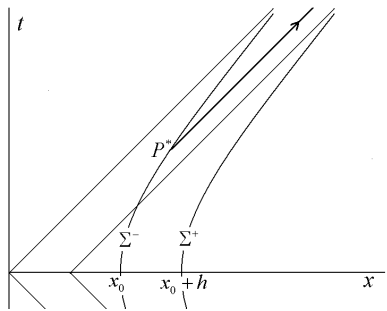}
\caption[short TOC caption]{\em Light ray emitted from $\Sigma^-$ in $P^*$ will
never reach $\Sigma^+$}
 \label{fig.never}
\end{center}
\end{figure}
but in this case the radar distance between the two hyperbolae is not constant.
There exist also paradoxical situations, as that exhibited in Fig.~\ref
{fig.never}, where $\Sigma^-$ cannot send a light signal that reaches $\Sigma
^+$. From this fact one should argue that it is not possible to treat a uniform
gravitational fields with an extended FFRF. But before giving up we will fully
investigate physical properties of extended frames which see $\Sigma^+$ and
$\Sigma^-$ describing Rindler hyperbolae of Eq.~(\ref{eq.rin2}) and
Eq.~(\ref{eq.rioo}). We will call such frames Rindler FFRF (and Rindler's field
those generating such kind of free fall). This kind of FFRF is the only one
that satisfy three important physical experimentally verified conditions:
\begin{itemize}
\item[(1)] Local acceleration does not depend on time;
\item[(2)] The ratio between proper time intervals at different altitudes is
also constant in time (possibly depending on the frames altitude), otherwise a
Pound and Rebka-like experiment would give every day a different result;
\item[(3)] Radar distance between two observers located at a different height
is a constant quantity which is time independent (it will as well possibly
depend on the altitude).
\end{itemize}
Property (1) is surely satisfied because it is the basis used for deriving
Rindler motion equation \cite{rindler}. Property (2) and (3) were verified by
Desloge \cite{des2} during the analysis of a uniform accelerated frame
(described with Rindler hyperbolae). In the next subsection we will prove them
in a way that appear to be more intuitive and probably more pedagogical.
\subsection{Lorentz invariance}
\label{lorentz}
We required the FFRF to be extended as in SR, therefore it must satisfy
Lorentz invariance. For proving it, let us apply a boost $\Lambda$ of velocity
$v= arcth \theta$ along $x$ direction to the hyperbola parametric equation
\cite{rindler,misner,mould} describing $\Sigma^-$
\be
\label{eq.th}
\mathbf {\Lambda}= \left(\begin{array}{ccc}
\cosh\theta  & -\sinh\theta  \\[5pt]
-\sinh\theta  & \cosh\theta
\end{array}\right)
\hspace{2cm}
\begin{cases}
&x=x_0\cosh g \tau\\
&t=x_0\sinh g \tau
\end{cases}.
\ee
The transformed hyperbola
\be \label {eq.parametricboost}
\begin{cases}
&x'=x_0\cosh (g \tau-\theta)\\
&t'=x_0\sinh (g \tau-\theta)
 \end{cases},
\ee
shows that the only effect of a Lorentz boost is a shift in the proper time
origin $\Delta \tau=-\theta/g$.
\par The application of Lorentz boosts on hyperbolae leaves their shape invariant
and brings on the $x$ axis all events having the same velocity $v= arcth
\theta$. This proves Lorentz invariance of Rindler FFRF and shows that, for
every given coordinate velocity $V=x/t$ (straight lines through the origin)
there exist a Rindler FFRF in which the events on this line are simultaneous.
\subsection{Proper time intervals}
\label{proper}
We need to calculate the proper time interval along a world line between two
generic coordinate times $t_1$ and $t_2$ and using units with $c=1$ we have the
usual expression $\tau_{12} = \int_{t_1}^{t_2}\sqrt{1-V^{2}}dt$.
Differentiating Eq.~(\ref{eq.rin2}) or Eq.~(\ref{eq.rioo}) one obtains $V =t/x$
and using Eq.~(\ref{eq.rin2}) and Eq.~(\ref{eq.rioo}) in order to eliminate
$x$, we can calculate the proper time interval $\tau^-$ along the lower and
$\tau^+$ along the upper hyperbola
\be \label{eq.tau0} \tau^-_{12} =
\int_{t_1^-}^{t_2^-}\frac{1}{\sqrt{1+(\frac{t}{x_0})^2}}dt
=x_0\int_{\frac{t_1^-}{x_{0}}}^{\frac{t_2^-}{x_{0}}}\frac{1}{\sqrt{1+y^{2}}}dy
\ee
\be \label{eq.tau+} \tau^+_{12} =\int_{t_1^+}^{t_2^+}\frac{1}
{\sqrt{1+(\frac{t}{x_0+h})^2}}dt
=(x_0+h)\int_{\frac{t_1^+}{x_{0}+h}}^{\frac{t_2^+}{x_{0}+h}}
\frac{1}{\sqrt{1+y^{2}}}dy. \ee
\begin{figure}[hbp]
\begin{center}
\includegraphics[scale=0.23]{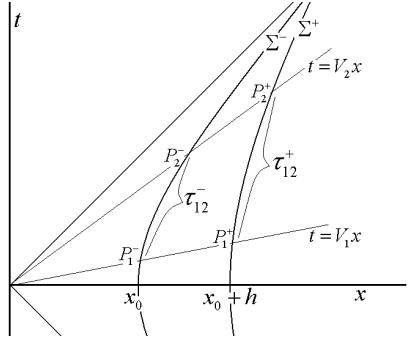}
%0.28
\caption[short TOC caption]{\em Proper times in $\Sigma^+$ and $\Sigma^-$}
\label{fig.clockdelay}
\end{center}
\end{figure}
Considering the intersection of the two hyperbolae with two lines of constant
coordinate velocity (see Fig.~\ref {fig.clockdelay})
\be t=V_1 x \qquad t=V_2 x \qquad \qquad |V_1|,|V_2|<1, \ee
one obtains two couples of points $P=P(x,t)$ in the Minkowski diagram, whose
coordinates are:
\be \label{eq.point}
P_1^-= \frac{\gamma _1}{g^-}\big (1,V_1 \big); \quad
P_1^+=\frac{\gamma_1}{g^+} \big (1,V_1 \big ); \qquad P_2^-= \frac{\gamma
_2}{g^-}\big (1,V_2 \big); \quad P_2^+= \frac{\gamma_2}{g^+} \big (1,V_2 \big
), \ee
where we have introduced the usual notation $\gamma=1/ \sqrt {1-V^2}$.
Coordinates of $P^-$ and $P^+$ are in simple proportions, and we have in
particular
\be \frac {t_1^+}{t_1^-}=\frac {t_2^+}{t_2^-}= \frac{g^-}{g^+}=\frac
{x_0+h}{x_0}, \ee
that proves the equality of integration extremes defining $\tau_{12}^+$ and
$\tau_{12}^-$. Dividing Eq.~(\ref{eq.tau+}) by Eq.~(\ref{eq.tau0}), one finds:
\be \label{eq.rap} \frac {\tau_{12}^+}{\tau_{12}^-} =\frac{g^-}{
g^+}=\frac{x_0+h}{ x_0}, \ee
so that one can conclude that:
\begin{quote}
{\em Proper times intervals along two Rindler hyperbolae between two events
having the same coordinate velocity, are in a fixed proportion.}
\end{quote}
We shown that all events having the same coordinate velocity $V=x/t$ can be
regard as simultaneous, and for this reason the above statement will prove
property (2).

\par Considering that $x_0=1/g^-$, Eq.~(\ref {eq.rap}) can also be
recast in the following form:
\be \frac {\Delta \tau}{\tau}=\frac
{\tau_{12}^+-\tau_{12}^-}{\tau^-_{12}}=g^-h, \ee
which was proved by Desloge during his analysis of uniform accelerated
reference frames in SR \cite{des3,des4}.
\subsection{Radar distance and light signals between Rindler hyperbolae}
\label{light}
\begin{figure}[hbp]
\begin{center}
\includegraphics[scale=0.18 ]{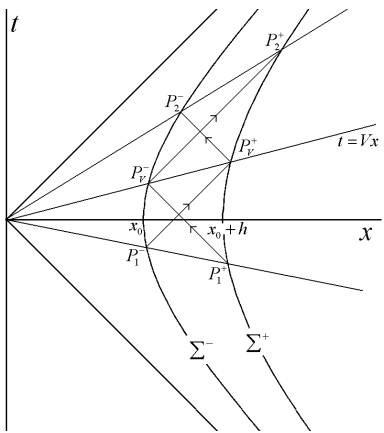}
\caption[short TOC caption]{\em Light signal between two hyperbolae}
\label{fig.aligned}
\end{center}
\end{figure}

Let us draw a generic line through the origin in Minkowski space, which
identifies points having the same coordinate velocity $V=x/t$:
\be t=Vx \qquad \qquad  |V|<1. \ee
Intersecting this line with the low and upper hyperbolas identifies the points
\be P_V^-=\frac{ \gamma}{g^-} \big (1,V \big);\qquad \quad P_V^+=
\frac{\gamma}{g^+} \big (1,V \big). \ee
The equations of light rays departing from $P_V^-$ and $P_V^+$ are respectively
\be t=x-\gamma \frac {1-V}{g^-}; \qquad t=-x+\gamma \frac{1+V}{g^+}. \ee
Solving the system together with the hyperbola equations (\ref{eq.rin2}) and
(\ref{eq.rioo})
\be
\begin{cases}
&t=x-\gamma \frac {1-V}{g^-}\\[5pt]
&x^{2}-t^{2}= \frac{1}{(g^+)^2}
\end{cases}
\hspace{2cm}
\begin{cases}
&t=-x+\gamma \frac{1+V}{g^+}  \\[5pt]
&x^{2}-t^{2}=\frac{1}{(g^-)^{2}}
\end{cases},
\ee
leads to the following intersection points:
\be
 P_2^- \!\! = \!\! \frac{\gamma g^+}{2} \!
\Big ( \! \frac{1+V}{(g^+)^2} \! + \! \frac{1-V}{(g^-)^2}, \frac{1+V}{(g^+)^2}
\! - \! \frac{1-V}{(g^-)^2} \! \Big ); \quad
 P^+_2 \!\! = \!\! \frac{\gamma g^-}{2} \!
\Big ( \! \frac{1+V}{(g^+)^2} \! + \! \frac{1-V}{(g^-)^2}, \frac{1+V}{(g^+)^2}
\! - \! \frac{1-V}{(g^-)^2} \! \Big ). \ee
In the same way it is possible to calculate intersection points between the two
hyperbolas and the incoming light rays:
\be
 P_1^- \!\! = \!\! \frac{\gamma g^+}{2} \!
\Big ( \! \frac{1+V}{(g^-)^2} \! + \! \frac{1-V}{(g^+)^2} , \frac{1+V}{(g^-)^2}
\! - \! \frac{1-V}{(g^+)^2} \! \Big ); \quad P^+_1 \!\! = \!\! \frac{\gamma
g^-}{2} \! \Big ( \! \frac{1+V}{(g^-)^2} \! + \! \frac{1-V}{(g^+)^2},
\frac{1+V}{(g^-)^2} \! - \! \frac{1-V}{(g^+)^2} \! \Big ). \ee
For any value of $V$, we have the following proportions
\be \frac{x_1^+}{x_1^-}=\frac{x_2^+}{x_2^-}=\frac{t_1^+}{t_1^-}=
\frac{t_2^+}{t_2^-}=\frac{g^-}{g^+}=\frac{x_0+h}{x_0}, \ee
from which one see that $P_1^-$ is aligned with $P^+_1$, and $P_2^-$ is aligned
with $P^+_2$ (as shown in Fig.~\ref {fig.aligned}).
We have thus found the following result which, as far as I know, was not
previously explicitly  pointed out:
\begin{quote}
{\em Light rays through a couple of events with the same coordinate velocity
laying on two Rindler hyperbolae, intersect other Rindler hyperbolae in a
couple of points that have in turn the same coordinate velocity}.
\end{quote}
This property it is an extra bonus, as we did not request it in our premise at
the beginning of section \ref {acc}, and it will make it easier to perform
clock rate comparison in an operational way.
Using Lorentz transformation we proved that all events having the same
coordinate velocity admit a Rindler FFRF in which they are simultaneous. As a
result of the above calculations, this appear now more clear, when one imagines
two observers in $\Sigma^-$ and in $\Sigma^+$ exchanging light signals.

\par We can now calculate the proper time $\tau_S^-$ elapsed in $\Sigma^-$
when a light ray covers the distance necessary to go from $\Sigma^-$ to $\Sigma
^+$ and the proper time $\tau_R^-$ taken for the return trip:
\be \tau_S^-\!\!
=\!\!\int_{t^-_V}^{t^-_2}\!\!\frac{1}{\sqrt{1+(\frac{t}{x_0})^2}}dt=
x_0\!\!\int_{\frac{t^-_V}{x_{0}}}^{\frac{t^-_2}{x_{0}}}\!\!\frac{1}{\sqrt{1+y^{2}}}dy;
\qquad
\tau_R^-\!\!=\!\!\int_{t^-_1}^{t^-_V}\!\!\frac{1}{\sqrt{1+(\frac{t}{x_0})^2}}dt=
x_0\!\!\int_{\frac{t^-_1}{x_{0}}}^{\frac{t^-_V}{x_{0}}}\!\!\frac{1}{\sqrt{1+y^{2}}}dy.
\ee
The calculation leads to constant quantities \footnote {If we tried instead to
calculate the coordinate $t$ time elapsed during the trip between $\Sigma^-$
and $\Sigma^+$, we would find the following expression $ t=\gamma \frac
{h(2x_0+h)}{x_0+h}$, which is not constant due to its dependence on $\gamma$.}

\be \label{eq.radar-} h^-=\tau_S^-=\tau_R^-=x_0\ln{\frac{x_0+h}{x_0}}=\frac
{1}{g^-}\ln {(1+g^-h)}. \ee
Calculation of proper times relative to an observer at rest with $\Sigma^+$
leads also to an equal and constant time, whose value is in this case:
\be \label{eq.radar+} h^+=\tau ^+_S=\tau^+_R= (x_0+h)\ln
{\frac{x_0+h}{x_0}}=\frac {1}{g^+}\ln {(1+g^-h)}. \ee
This is an important result that, differently from Desloge \cite{des2}, we got
by direct calculation:
\begin{quote}
{\em The proper time, as measured by an observer on a Rindler hyperbola, taken
by a light pulse to reach another Rindler hyperbola is equal to the proper time
taken for the return trip, and this proper time is constant.}
\end{quote}
With the proof of this property, we completely fulfilled request (3) of section
\ref{acc}.
\par In Eq.~(\ref {eq.radar-}) and Eq.~(\ref{eq.radar+}) we calculated the proper time
elapsed in $\Sigma^-$ and $\Sigma^+$ for sending and receiving a light ray.
Multiplying this time by the speed of light ($c=1$ in our units) we have radar
distances (we used $h^-$ for radar distance measured by $\Sigma^-$ and $h^+$
for that measured by $\Sigma^+$).
\section{Physical measurements}
\label{measures}
\subsection{Clock delay}
\label{mis}
All necessary physical requirements having thus been satisfied, we feel
entitled to set up a thought experiment in which two identically constructed
clocks sit at rest, in a uniform and static gravitational field, at fixed
distances along a field line of force. As has become customary, we will give a
name to the physicists involved in the measuring processes; let's say that
Alice enjoys her life staying at rest with $\Sigma^-$ and Bob staying higher in
$\Sigma ^+$. They will perform a measure in the following way (see
Fig.~\ref{fig.aliceBob}):
\begin{figure}[hbp]
\begin{center}
\includegraphics[scale=0.20 ]{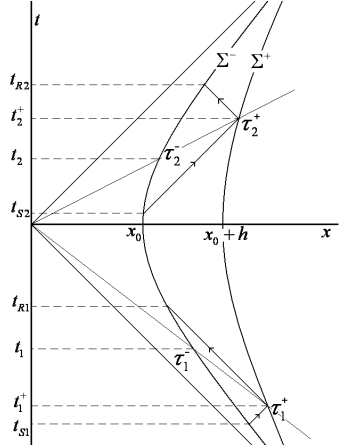}
\caption[short TOC caption]{\em Alice and Bob compare their clock rate}
\label{fig.aliceBob}
\end{center}
\end{figure}
\begin{description}
\item[a)] at time $t_{S1}$ Alice sends a light signal upwards;
\item[b)] at time $t^+_1$ Alice's light pulse is received by Bob. When Bob
receives the signal he starts his clock and immediately sends back downwards a
return light signal;
\item[c)] at time $t_{R1}$ Alice receives the return signal from Bob. Adopting
Einstein clock synchronization, she will argue that Bob clock started at the
intermediate time between sending and receiving the light pulse. According to
her clock this time is $\tau_1^-=\frac{\tau_{S1}^-+\tau_{R1}^-}{2}$, and
corresponds to coordinate time $t_1$. In the previous section we saw that, with
respect to the origin, this event is aligned with the event in which Bob
started his clock;
\item[d)] at time $t_{S2}$ Alice sends a light signal upwards;
\item[e)] at time $t^+_2$ Alice light pulse is received by Bob. When Bob
receives the signal he stops his clock and immediately sends back to Alice a
return signal, in which the information of the proper time elapsed in Bob's
clock is stored;
\item[f)] at time $t_{R2}$ Alice receives the return signal from Bob. She will
argue that Bob clock had stopped at the intermediate time between sending and
receiving the light pulse. According to her clock this time is
$\tau_2^-=\frac{\tau_{S2}^-+\tau_{R2}^-}{2}$ which correspond to coordinate
time $t_2$.  We have again that in the Minkowski space-time of $K$, this event
is aligned with the origin and the event in which Bob stopped his clock.
\end{description}
The necessary measurements have thus been done, and Alice can compare her time
with Bob's. It is immediately seen that any dependence on starting and stopping
times cancels out, and that she will be able to compare two proper times seen
under the same angle centered in the origin: $\tau^-_{12}=\tau^-_2-\tau^-_1$;
$\tau^+_{12}=\tau^+_2-\tau^+_1$.
These proper times satisfy Eq.~(\ref{eq.rap}) and Alice concludes\footnote{The
experiment could be also done by sending starting and stopping signals from
$\Sigma^+$. In this case Bob would draw the very same conclusion as Alice and
they would both agree on the fact that Bob's clock is faster than Alice's. It
worth stressing that this situation is different from that arising in SR with
respect the time dilation effect, which is observed in a symmetric way by the
two observers.}
\be \label{eq.dilat} \frac{\Delta
\tau}{\tau}=\frac{\tau^+_{12}-\tau^-_{12}}{\tau^-_{12}}=g^-{h}. \ee
This fact was proved by Desloge \cite{des4} for uniform accelerated frames, but without
having recourse to an ideal experiment as performed in this paper .
It is also possible to compare Bob's proper time $\tau^+_{12}$ with Alice's
calculated either between the two emission or reception events. This would not
make any difference, because in both cases we would have only added to and then
subtracted from $\tau^-_{12}$ the same quantity (the times taken by the outward
and return trips given by $\tau _S^- = \tau _R^- = x_0 \ln{(1+g^-h)}$).
\subsection{Space dilation}
\label{spacedil}
Solving Eq.~(\ref{eq.radar-}) and Eq.~(\ref{eq.radar+}) for $h$ one gets
\be \label{eq.inverseradar} h=x_0(e^{g^-h^-}-1)=x_0(e^{g^+h^+}-1), \ee
which makes it possible to rewrite the time dilation formula using local radar
coordinates:
\be \label{eq.rinexp} \frac{\tau^+}{\tau^-}=1+g^-h=e^{g^-h^-}=e^{g^+h^+}, \ee
result pointed out from Desloge \cite{des4}, in the case of uniform accelerated
frames. It is also interesting to prove this result following redshift derivations based
on potential energy loss of a quantum particle \cite{fabri,misner2,schutz}
\footnote{Let us imagine a particle of mass $m_0$ falling from $\Sigma^+$ to
$\Sigma^-$. From the FFRF's viewpoint, the falling mass will arrive at $\Sigma^-$
with total energy $E^-=m_0(1+g^-h)$. Once in $\Sigma^-$, the particle transforms
into a photon `climbing' to $\Sigma^+$. The related emitted frequency
(in unit with $h_{plank}=1$) is $\nu_e=m_0(1+g^-h)$ while (for the conservation law)
the received frequency in $\Sigma^+$must be again $\nu_r=m_0$. As expected,
the frequency ratio is:
\be \nu_e/\nu_r=1+g^-h=e^{g^+h^+}=e^{g^-h^-}.\ee
Using Eq.(\ref{eq.grel}) it easy to show that also an observer comoving with
the FFRF in $x_0+h$ gets the same expression. If an observer sitting $\Sigma^+$ and
$\Sigma^+$ used the same calculation procedure, he would instead find a wrong expression
($\nu_e/\nu_r=1+g^-h^-=1+g^+h^+$).}.
\par One can prove that Eq.~(\ref{eq.rinexp}) is
consistent with the following general result of GR \cite{rindler2},
\be \label{eq.rinorig} \frac {\nu^+}{\nu^-}=e^{ \phi(x_0+h)-\phi(x_0)}, \ee
which expresses the redshift in terms of the Newtonian potential $\phi$ of the
gravitational field. Indeed, calculating the potential difference we find:
\be \phi (x_0+h)-\phi (x_0)=\int_{x_0}^{x_0+h}g(x)dx =\ln \frac{x_0+h}{x_0}=\ln
(1+g^-h) \ee
which, inserted in Eq.~(\ref{eq.rinorig}), reproduces the time dilation formula
(\ref{eq.rinexp}).
For a uniform gravitational field Desloge \cite{des4} proved instead the expression
$\frac{\nu^+}{\nu^-}=e^{gh}$, formally equivalent to Eq.~(\ref{eq.rinexp}) but
here the gravitational acceleration is the same for top and bottom observers and
the distance is measured from a FFRF. Its derivation is based on the following
metric \cite{des4}:
\be \label{eq.metricaexpo} ds^2=e^{2gh}dt^2-dh^2, \ee
that is obtained requiring the FFRF to be rigid with its parts moving on
geodetics with constant acceleration. Desloge's metric of
Eq.~(\ref{eq.metricaexpo}) is associated to a curved space with non vanishing
Riemann tensor ($R_{xttx}=g^2e^{2gh}$) and Ricci tensor ($R_{tt}=g^2e^{2gx}$,
$R_{xx}=-g^2$) thus,  for the Einstein field equation, the associated space is
not empty. Because of the curvature associated to this metric, it is also not
possible to have an extended FFRF. In the next section we will see that
Eq.~(\ref{eq.rinexp}) is instead consistent with the metric derived from
Einstein equation requiring the space to be flat ($R_{\mu \nu \sigma \rho}=0$)
\cite{ror}.
\par For a better understanding of space dilation it is useful to calculate
it as measured with radar methods. Considering the ratio between Eq.~(\ref {eq.radar-})
and Eq.~(\ref{eq.radar+}) we get the following expression analog to time dilation:
\be \label{eq.distanceratio} \frac{h^+}{h^-}=\frac{x_0+h}{x_0}=\frac{g^-}{g_+}.
\ee
From this relation one find that the work per unit mass it is the same for
observers in $\Sigma^-$ and $\Sigma^+$ ($ g^-h^-=g^+h^+ $).
\par It is also interesting to calculate the expression of the infinitesimal
distance measured with radar methods by two observers at rest with the field at
different heights. We will use $dh^-_x$ for denoting radar measurements of an
infinitesimal length located at coordinate $x$, performed by an observer
sitting in $\Sigma^-$, and $dh^+_x$ if the observer is located at $\Sigma^+$. A
straightforward calculation, at first order, gives:
\be \label {eq.delta} dh^-_x\!=\!x_0\! \bigg(\! \ln{\frac{x+dh}{x_0}}-\ln
{\frac{x}{x_0}}\!\bigg) \!\approx \!\frac{x_0}{x}dh;\qquad dh^+_x\!=\!(x_0 +
h)\! \bigg (\!\ln{\frac{x+dh}{x_0}}  - \ln {\frac{x}{x_0}} \! \bigg) \!\approx
\!\frac{x_0+h}{x}dh, \ee
whose ratio is:
\be \label{eq.deltah} \frac {dh^+_x}{dh^-_x}=\frac{x_0+h}{x_0}=\frac{g^-}{g^+},
\ee
from which one sees also that $\Sigma^-$ and $\Sigma^+$ measure the same
infinitesimal work per unit mass: $d \phi =g^+dh^+=g^-dh^-$.
\subsection{Gravitational acceleration}
\label{gravityacc}
In this subsection, we want to compare physical measurements performed by
different observers in the same local frame. Let us consider two different
measurements in Bob's frame ($\Sigma^+$):
\begin{description}
\item [1] Bob ($\Sigma^+$) performs direct measurements with clocks and rods
built in $\Sigma^-$

\item [2] Alice ($\Sigma^-$) observes Bob's frame by sending and receiving
reflected light pulses.
\end{description}
The two measurements are in the following relations
\be \label{eq.proportion}
\underbrace{dh_{x_0+h}^-}_{Alice's~measure}=\frac{g^+}{g^-}
\underbrace{dh_{x_0+h}^+}_{Bob's~measure};\qquad \quad
\underbrace{d\tau_{x_0+h}^-}_{Alice's~measure}=\frac{g^+}{g^-}
\underbrace{d\tau_{x_0+h}^+}_{Bob's~measure}. \ee
Therefore, for the same couple of events, Alice obtains the same average
velocity measured by Bob, but measuring shorter space and time intervals.
Because of these contractions, it easy to prove that Alice will measure
larger accelerations than those measured by Bob:
\be \label{eq.gratio} \underbrace{g_{x_0+h}}_{Alice's~measure}=
\frac{g^-}{g^+}\underbrace{g_{x_0+h}}_{Bob's~measure}. \ee
We know that the value of the gravitational acceleration obtained by Bob is
$g^+$ and, from Eq.~(\ref{eq.gratio}), we learn that Alice will find that the
gravitational acceleration in Bob's frame is $g^-$. It is easy to conclude
that:
\begin{quote}
{\em Every observer sitting at rest with the source of gravity in a Rindler
field finds that the gravitational acceleration has the same value everywhere.}
\end{quote}
\par
\begin{figure}[hbp]
\begin{center}
\includegraphics[scale=0.22 ]{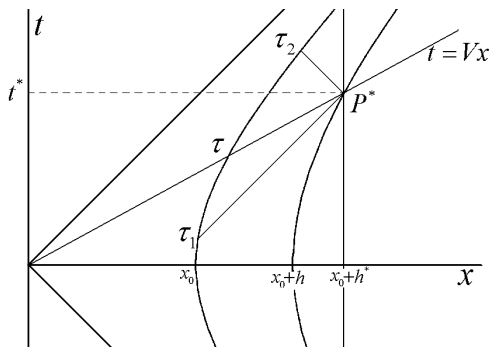}
\caption[short TOC caption]{\em Radar measure of a free falling body performed
by $\Sigma^-$ } \label{fig.appearuniform}
\end{center}
\end{figure}

 For a more exact treatment we will now calculate the position of a body at rest with
the FFRF as seen by radar measurements from an observer located in $\Sigma^-$.
One must calculate the proper time needed to go (or come back) from $\Sigma^-$
to a fixed coordinate distance $h^*$ as a function of the intermediate proper
time $\tau=\frac{\tau_S+\tau_R}{2}$ (that we proved to be the time elapsed in
$\Sigma^-$ to arrive at the point of intersection with the straight line $t=Vx$
through $P^*$, as shown in Fig.~\ref{fig.appearuniform}).
It is possible to verify that $\Sigma^-$ proper time is
\be \label{eq.tau}
\tau^-=\frac{x_0}{2}\ln{\frac{1+V}{1-V}}. \ee
Inverting this relation (or directly from Eq.~(\ref{eq.th})), using $g$ for
denoting $g^-$, and $\tau$ for $\tau^-$, one finds
\be \label{eq.coordvel} V=\tanh{g\tau}. \ee
Using Eq.~(\ref{eq.point}) one finds that at $t=0$ the $x$ coordinate of the
hyperbola passing for $P=(x_0+h^*,t^*)$ is $x_0+h=(x_0+h^*)\sqrt{1-V^2}$.
Putting this value in Eq.~(\ref{eq.radar-}), one can calculate the radar
distance $h_{rad}$ from $\Sigma^-$ of a body falling from the initial coordinate
$h^*$ as a function of the coordinate velocity $V$:
\be \label{eq.radargamma}
h_{rad}(V)=x_0\ln{(1+gh^*)}+\frac{x_0}{2}ln{(1-V^2)}, \ee
and, using Eq.~(\ref{eq.coordvel}) to eliminate $V$, one gets the following motion equation:
\be \label {eq.radarposition}
h_{rad}(\tau)=h_{rad}(0)-x_0\ln({\cosh{g\tau}}). \ee
Deriving with respect to $\Sigma^-$ proper time $\tau$, one finds the corresponding
velocity $v_{rad}(\tau)=-\tanh{g\tau}$ which has the same value and opposite
direction of the coordinate velocity $V(\tau)$ and does not depend on the
starting point of the fall $h^*$, but only on $\Sigma^-$ acceleration $g=1/x_0$.
For this reason, an observer at rest with the field observes every part of the
FFRF to move at the same speed.
\par One can also calculate the acceleration $a_{rad}(\tau)=-g\cosh^{-2}{g\tau}$,
whose value does not depend on $h^*$. For $\tau=0$ we have $a_{rad}(0)=-g$
and this confirms the fact (qualitatively discussed above) that, from the viewpoint of
an observer sitting in $\Sigma^-$, the field under examination appears to be
spatially uniform. In contrast, a falling body does not appear to move with an
uniform proper acceleration, because $\tau$ it is not the proper time of the
observed falling body.
The same reasoning applies to an observer located in $\Sigma^+$, but he observes
everywhere a uniform acceleration with a smaller intensity $g^+$.
\par It is also interesting to note another peculiarity of the Rindler's field:
moving along the $x$ direction in a infinite Rindler's field, the local gravitational
acceleration $g$ will assume every possible value. This has an important
consequence:
\begin{quote}
{\em All infinite extended Rindler's fields are equivalent.}
\end{quote}
This seems a strange conclusion, but remember that our starting point was
purely ideal, and we did not analyse whether and how such a field could be
realized in the real world. However, some papers \cite{ava,tex} discusses a
possible connection of the Rindler's metric (which is found as a particular case of
the Taub metric) with the study of the exterior solution of a planar symmetry mass
distribution. If this will be confirmed, one would find that the field
strength would not depend on the mass value, and that the field would not
vanish even with a vanishing mass. Desloge's metric of Eq.~(\ref{eq.metricaexpo})
is instead not compatible with a planar mass distribution as, because of the
Einstein field equation, it is associated to an non-empty space even out of the
plane. As far as I know, most of the considerations presented in this sections
have not benn pointed out before.
\section{Metric Properties}
\label{metric}
In section \ref{measures} we calculated the time ratio between two clocks at
rest with the field $\frac {\tau}{\tau_0}=1+gx$, where now, in order to denote
the height measured from a FFRF (which starts the fall at $t=0$) we use $x$
instead of $h$, and $g$ instead of $g^-$. Since the field we consider here is
static, it is possible to write the metric in diagonal form
\be \label{eq.metric} ds^2=A(x)dt^2-B(x)dx^2. \ee
Writing the proper time elapsed for a clock at rest with the field
$d\tau=\sqrt{A(x)}dt$, comparing $d\tau$ with $d\tau_0$ is easy to prove that
the explicit expression of the first metric term is $A(x)=(1+gx)^2$.
In order to find that of $B(x)$, it is possible to relate radar distances
expressed using the metric coefficients with that we calculated in
Eq.~(\ref{eq.radar-}). One has to solve the equation
\be \label{eq.radarandmetric}
A(0)\int_{0}^{x}{\sqrt{\frac{B(k)}{A(k)}}dk}=\frac{1}{g}\log{(1+gx)}. \ee
Setting $A(x)=1+gx$ and differentiating one finds:
\be \frac{\sqrt{B(x)}}{1+gx}=\frac{1}{1+gx}, \ee
hence $B(x)=1$. The above result can also be deduced using the general flat
space solutions of the Einstein field equation (imposing the condition
$R_{\mu\nu \sigma \rho}=0$). Rohrlich was the first to show that in this case
we have the following relation between the metric coefficients \cite{ror}
\be \label{eq.ror} B(x)=\bigg ( \frac{1}{g}\frac{d}{dx}\sqrt{A(x)}\bigg )^2 \ee
and, for $A(x)=(1+gx)^2$,  we find again $B(x)=1$. The resulting metric line
element
\be \label{eq.real} ds^2=(1+gx)^2dt^2-dx^2 \ee
is deduced also in Mould's treatise \cite{mould}, in a formally different way,
also based on EP and on GR formalism (without involving the field equation), and
its coordinate are relative to the FFRF.
The Rinlder metric of Eq.~(\ref{eq.real}) is only one particular solution of the field
equation, and it worths underlying that, without having discussed the direct EP
application, it could not be possible to choose this solution between the
infinity that satisfy Eq.~(\ref{eq.ror}). Rohrlich stressed that between these
solutions three of them have a particular physical interest \cite{ror}. The
first one gives Eq.~(\ref{eq.real}), while the other two give the following
metrics:
\be \label{eq.metric3} ds^2=(1+2gx')dt'^2-(1+2gx')^{-1}dx'^2 \ee
\be \label{eq.metric2} ds^2=e^{2gx''}(dt''^2-dx''^2), \ee
which are easily obtained from Eq.~({\ref{eq.real}) using the following
coordinate transformations
\be \label{eq.trasfgut} 1+gx=(1+2gx')^{1/2} \qquad t=t' \ee
\be \label{eq.trasfdes} 1+gx=e^{gx''} \qquad t=t''. \ee
Analysing the transformation of Eq.~(\ref{eq.trasfdes}) it easy to check that
spatial coordinates used in the metric of Eq.~(\ref{eq.metric2}) are relative
to an observer at rest with the field. This metric was first deduced from Lass
\cite{lass} in the study of uniform accelerated observers.
\par Let us compare the above metrics with the Schwarzschild line
element (dropping the angular term)
\be \label{eq.Schwarz1} ds^2=\bigg (1-\frac{2GM}{r}\bigg )dt^2-\bigg
(1-\frac{2GM}{r}\bigg )^{-1}dr^2. \ee
\par The Taylor expansion at second order gives
\be \label{eq.Schwarzapp} A(x)=B^{-1}(x)=\bigg (1-\frac{2GM}{r}\bigg
)\approx\bigg (1-\frac{2GM}{r_0}
+\frac{2GM}{r_0^2}(r-r_0)-\frac{2GM}{r_0^3}(r-r_0)^2\bigg ). \ee
With the substitution $x=(r-r_0)$, and $g=GM/r_0^2$, one obtain the following
metric
\be \label{eq.Schwarz2}
ds^2=(1-2gr_0+2gx-2\frac{g}{r_0}x^2)dt^2-(1-2gr_0+2gx-2\frac{g}{r_0}x^2)^{-1}dx^2.\ee
If the condition $|x|<<r_0$ it is satisfied, one can take only first order
terms, and translating the spatial origin ($x'=x-r_0$) one obtains
Eq.~(\ref{eq.metric3}). This approximation must be used with care, having lost some important
properties of the Schwarzschild solution, the most important of those is the
Riemann curvature tensor $R_{\mu \nu \sigma \rho}$ that now has become zero
everywhere (in the Schwarzschild solution $R_{ \mu \nu}=0$, but $R_{\mu \nu
\sigma \rho} \neq 0$). But this metric describes the low gravity limit, that is
the only place in the real world where one can find a good approximation of a
uniform gravitational field. Metric in Eq.~(\ref{eq.metric3}) is also called
`Nearly Newtonian Gravitational Field' \cite{misner3}.
\par Starting instead from Desloge' metric of Eq.~(\ref{eq.metricaexpo}) and operating
the transformation $e^{gx}=1+gx'$, one gets
\be \label{eq.deslogequad} ds^2=(1+gx')^2dt^2-(1+gx')^{-2}dx^2. \ee
This metric does not reproduce the Nearly Newtonian Gravitational Field of
Eq.~(\ref{eq.metric3}), and its second order terms are different from those of
the Schwarzschild expansion of Eq.~(\ref{eq.Schwarz2}).
\section{A derivation of Desloge's metric}
\label{desloge}
In deriving the metric of Eq.~(\ref{eq.metricaexpo}) Desloge \cite{des3} states
that it is necessary to set $B(x)=1$ to preserve the rigidity of the FFRF,
but he does not discuss this point in detail.
In our approach to the EP we derived also $B(x)=1$, but with a different temporal
metric coefficient $A(x)$. In this section we want to prove instead that it does exist a possible
approach based on Rindler's hyperbolae that brings to Desloge's metric, but it
requires argumentations not contained in the EP. The `different' physics
consists in supposing that the RF falling with half of the gravitational
acceleration, $g'=g/2$, is inertial. An observer comoving with this frame (denoted
with $D$) would see both the `free fall' frame $K$ and the rest frame $\Sigma$
describing Rindler hyperbolae, as illustrated in Fig.~\ref{fig.desloge}
\footnote{For simplicity, we drew $K$ and $\Sigma$ trajectories in separate
planes, but the correct physical interpretation holds when one translate the
two RF so that $K^0$ and $\Sigma^0$ touch at $t=0$. The coordinates of $K$ are
in fact physically positive and the negative sign is used only for preserving
the correct relation between coordinate and acceleration ($x=1/g$). With this
particular approach we have also that, from the point of view of $D$, the force
accelerating $K$ appear now to increase with the altitude. Another new feature
it is that there exist now two points (at finite distance) with infinite
acceleration. We remark also that this hyperbolae displacement it is not Lorentz
invariance and now, for this reason, the frames can have only local validity.}
\begin{figure}[hbp]
\begin{center}
\includegraphics[scale=0.22 ]{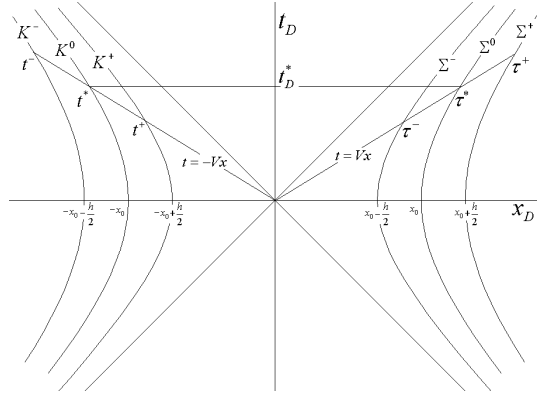}
\caption[short TOC caption]{\em Desloge's metric derivation using an intermediate RF}
\label{fig.desloge}
\end{center}
\end{figure}
\par From $D$'s viewpoint, at any fixed time $t^*_D$, $\Sigma^0$ and $K^0$ clocks ticks
the same proper time $t^*=\tau^*$ (they describe in fact the same trajectory with
respect to the inertial frame $D$). Since we proved that proper times of events having
the same coordinate velocity (on straight line $t=-Vx$ for $K$ and $ t=Vx$ for $\Sigma$)
are in a fixed proportion, we have:
\be \label {eq.ratios}
\frac{\tau^+}{\tau^-}=\frac{t^-}{t^+}=\frac{1+g'\frac{h}{2}}{1-g'\frac{h}{2}}.\ee
Sitting in the inertial frame $D$ and comparing the frequencies of falling clocks in
$K$ with those in the rest frame $\Sigma$, one finds that a clock in
$\Sigma^+$ runs faster than one in $K^+$, while a clock in $\Sigma^-$ ticks slower
than one in $K^-$, with a proportionality factor given by Eq.~(\ref{eq.ratios}).
For this two reasons, an observer living in $K$ observes clocks in $\Sigma^+$
running faster than those in $\Sigma^-$, with a proportionality factor that is
the square of that in Eq.~(\ref{eq.ratios}):
\be \label{eq.ratiok}
\frac{\tau_{\Sigma}^+}{\tau_{\Sigma}^-}=\bigg(\frac{1+g'\frac{h}{2}}{1-g'\frac{h}{2}}\bigg)^2.\ee
From Eq.~(\ref{eq.radar-}) one finds the radar distance $x$ between $K^+$ and $K^-$
that, in the view of an observer in $K^0$, is:
\be
x=\frac{1}{g'}\log{\bigg(\frac{1+g'\frac{h}{2}}{1-g'\frac{h}{2}}\bigg)}.
\ee
Putting the inverse of this expression in Eq.~(\ref{eq.ratiok}) one
obtains the time ratio of $\Sigma$ clocks in the radar coordinate of $K^0$
 \be
\tau_{\Sigma}^+/\tau_{\Sigma}^-=e^{2g'x}=e^{gx}.\ee
This proves that the metric coefficient is $A(x)=e^{2gx}$.
The spatial coefficient must be instead set to unit, $B(x)=1,$
because $\Sigma$ and $K$ do not show a relative contraction (they both contract
with respect to the inertial frame $D$), and one obtains Desloge's metric
$ ds^2=e^{2gx}dt^2-dx^2$.
\section{Concluding remarks}
\label{conc}
As far I know, part of the analysis presented here is new, and hopefully could
give some further physical insight on the problems at hand. My first intent in
writing this paper was mainly pedagogical. For this reason I have always tried to
use simple mathematics, and sometimes calculations are perhaps not performed in
the shortest way.
\par We conclude the discussion with some general considerations.
In the physical case of central Newtonian potentials it always exists an
asymptotic limit in which gravity vanishes, from where one can imagine to let
the fall of the FFRF start. This is used in GR to choose the correct form of
Schwarzschild metric, but unfortunately this cannot be achieved in the case of
uniform fields. Supposing the fall of an extended body to start from within a
region in which the field does not vanish, and taking into account the fact
that the light speed limit is valid for every signal, one finds large
differences depending on the way the body is released. For example, starting
the body fall removing a support from its bottom, one has a `dilation' effect,
because the top starts to fall later than the bottom, whereas releasing the
hook from which it hangs, the falling body will experience a `contraction'
during the fall. Rindler FFRF leads to a different rule: the fall starts
simultaneously with respect to every local frame  and it is `programmed' to
have the same velocity along lines through the origin; observers at rest with
the field will consider events on these lines simultaneous. Some of these
points are discussed in detail in Mould treatise \cite{mould}. Incidentally,
with this choice, we have found that the FFRF measure gravity acceleration
decreasing with height and this generate an asymptotic limit where gravity
vanishes. Unfortunately, this limit does not have a non-relativistic
(Newtonian) known counterpart, as it is in the case of Schwarzschild solution.
But because the Rindler's field is globally flat, one is free to choose
the origin $x_0$ where he like and to compare the acceleration with the
Newtonian one.
\begin{acknowledgments}
I am very grateful to Silvio Bergia who suggested me to study uniform gravitational
field using only accelerated free falling frames. He
gave me also moral support and useful suggestions during the preparation of
this work. Stimulating discussions with Corrado Appignani, Luca Fabbri, Mattia
Luzzi and Fabio Toscano are also gratefully acknowledged.
\end{acknowledgments}
\end{document}